\documentclass[aps,prd,twocolumn,10pt,reprint,nofootinbib]{revtex4-2}
\usepackage[table]{xcolor}
\definecolor{headerblue}{RGB}{38,70,83}
\definecolor{rowlight}{RGB}{245,248,252}
\definecolor{firstcol}{RGB}{233,240,248}
\definecolor{accent}{RGB}{42,157,143}
\definecolor{Gray}{gray}{0.85}
\definecolor{LightCyan}{rgb}{0.88,1,1}
\usepackage[colorlinks,linkcolor={red},citecolor={blue}]{hyperref}
\usepackage{eurosym}
\usepackage{graphicx}
\usepackage{amsmath}
\usepackage{amssymb}
\usepackage{tensor}
\usepackage{amsfonts}
\usepackage{bm}
\usepackage{hhline}
\usepackage{float}
\usepackage{makecell}
\usepackage{booktabs}
\usepackage{multirow}
\usepackage{rotating}
\usepackage{threeparttable}

\usepackage{multirow}

\def\be{\begin{equation}}
	\def\ee{\end{equation}}
\def\bea{\begin{eqnarray}}
	\def\eea{\end{eqnarray}}

\begin{document}
	\title{New generalization of the Barboza-Alcaniz parametrization of Dark energy}
	\author{Shahab Shahidi}
	\email{s.shahidi@du.ac.ir}
	\affiliation{School of Physics, Damghan University, Damghan 36716-45667, Iran.}
	\date{\today}
	
\begin{abstract}
A generalization of the Barboza-Alcaniz parametrization of dark energy is proposed. This is a three-parameter model which can resolve the shortcomings of the Barboza-Alcaniz behavior at future times. We show that cosmological data favor the new parametrization over both the Barboza-Alcaniz model and $\Lambda$CDM. We also consider the cosmological implications of the model and show that the qualitative behavior mimics to the original Barboza-Alcaniz model, with a slightly smaller acceleration rate.
\end{abstract}
\maketitle
	
\section{Introduction}
Cosmology has entered a new and interesting era. On one hand, we have a dozen of gravity theories, all of them claim to have a more complete insight/result than the standard model of cosmology, and on the other hand, we have a plethora of available datasets, from large galaxy surveys like DESI \cite{desi} or DES \cite{des} and Supernova light curve observations \cite{pantheon}, to precise measurements of the cosmic microwave radiation from Planck \cite{planck}, covering both local low to high $z$ ranges. 

Despite the large amount of cosmological data, all of them strongly support the accelerated expansion of the Universe, which can be simply and historically explained by adding a constant term to the Einstein's gravity theory. Together with the cold dark matter content which is needed to account for local physics, we reach to the $\Lambda$CDM model, where all observations has been successfully confronted.

However, there are some fundamental shortcomings related to the $\Lambda$CDM model which motivate cosmologists to think about an alternative. One of the main concerns about the $\Lambda$CDM model is the cosmological tensions, like the Hubble \cite{Hubbletension} and $S_8$ \cite{S8tension} tensions \footnote{The so-called $S_8$ tension refers to the difference of the value of the $S_8$ from weak gravitational lensing of KiDS \cite{kids} and those from the Planck data \cite{planck,cmb}. However, the updated analysis of KiDS has reduced the discrepancy to around $0.73\sigma$, seems to resolve the problem.}. Also the introduction of a cosmological constant in $\Lambda$CDM model rise a so-called cosmological constant problem, where the cosmological observations require an extremely small effective vacuum energy density compared to naive quantum field theory estimates by a factor of $\sim10^{-120}$ \cite{CCproblem}.

In order to solve (or at least shed light) to above concerns, it is necessary to generalize the cosmological constant in $\Lambda$CDM model to a dynamical field which can vary over time. This could be done easily by promoting the cosmological constant to a scalar field, adding a corresponding dynamical term and then a nonminimal coupling to gravity, to construct a scalar-tensor theory of gravity \cite{horndeski}. Further, one can construct a vector field theory \cite{vectortensor} or a multi-field modified gravity theory \cite{multifieldgravity}. 

Another indirect generalization of the cosmological constant could be achieved by generalizing a geometry part of the theory. This leads to $f(\textmd{geometry})$ models like $f(R)$ \cite{fR}, Weyl \cite{Weyl} or torsion-based models \cite{teleparallel}. From the field theory point of view, one can also modify the way the gravity interacts like the massive theories of gravity \cite{massivegravity}.

Independently, one can also modify the way the matter interact with itself and with geometry. Examples include $f(R,T)$, $f(R,L_m)$, $f(T_{\mu\nu}T^{\mu\nu})$, etc. \cite{modifiedmattermodels}. The main property of these kind of models is the non-conservation of the energy-momentum tensor, leading to creation of matter from geometry and vice versa and then produce an acceleration through the matter sector. A very recent attempt can be find in \cite{farahzad} where the acceleration is achieved through a generalized matter Lagrangian.

Despite that modifying gravity in the above sense are based on physical grounds and cosmologically motivated, there are always difficulties dealing with a modified gravity. Beside complexity, we are always encountered the stability and regularity issues. As a result, investigating the dark energy behavior of the Universe is more difficult in modified gravity. As a result, it would be very interesting to consider possible parametrizations of the dark energy (DE) which can be easily confronted with cosmological observations and directly reflect the behavior of the DE itself. There exist dozens of parameterizations of DE including the Chevallier, Polarski and Linderand (CPL) \cite{CPL}, Barboza, Alcaniz (BA) \cite{BA}, logarithmic, linear \cite{lieanrDE}, etc. The main concern of these parametrizations is to describe cosmological data with a small possible set of model parameters. Also, the behavior of the model at the boundaries of cosmic evolution and also at present time should also be take into account. It should also be noted that all of the DE parametrizations could be obtained from a specific modified gravity theory, but usually this is not an easy procedure.

In this paper, we will consider a new parametrization of the DE which could be seen as an extension of the BA parameterization.
DE parametrizations are not intended to infer the true future evolution of dark energy, but rather to construct a flexible, data-driven interpolation that remains well behaved near $z\leq 0$ and avoids the pathologies of divergent or ill-defined future behavior. In this work, we regard the proposed model as an observationally motivated phenomenological parametrization of the dark energy equation of state, rather than a microscopic description of its underlying physics. The main purpose is to provide a flexible bounded form that can be tested against cosmological data and compared with the BA and $\Lambda$CDM models. In this sense, the present study is directly connected to late-time cosmological observations and the phenomenology of cosmic acceleration.

In the next section, we will describe the model and also review in details the two most important parameterizations of the DE and their relations to this new model. In section \ref{sec3} we will show how this new parametrization could be obtained from a modified gravity theory. In sections \ref{sec4}, \ref{sec5} we will consider cosmological implications of the extended BA parameterization and compare to the $\Lambda$CDM and also the BA model. Section \ref{sec6} will be dedicated to conclusions and final discussions.

\section{The model} \label{sec2}
Let us assume that the Universe is homogeneous and isotropic, described by the FRW metric of the form
\begin{align}
		ds^2=-dt^2+a^2(dx^2+dy^2+dz^2),
\end{align}
where $a=a(t)$ is the scale factor and $t$ is the cosmic time.
We assume that the Universe is filled with a pressureless dust and a dark matter component with combined energy density $\rho_m$, radiation with energy density $\rho_r$ and pressure $P_r=1/3\rho_r$ and a dynamical dark energy (DE)  component with energy density $\rho_{de}$ and pressure $$p_{de}=w_{de}\rho_{de},$$
 where $w_{de}$ is a function of cosmic evolution. 
With these assumptions, one can write the Friedmann and Raychaudhuri equations as
\begin{align}
3H^2 &= 8\pi G \left( \rho_m + \rho_r + \rho_{de} \right),\\
3\left(\dot{H}+H^2\right) &= -4\pi G \left[ \rho_m + 2\rho_r + (1 + 3w_{de})\rho_{de} \right],
\end{align}
where $H$ is the Hubble parameter defined as $H=\dot{a}/a$ and $G$ is the gravitational constant. All matter fields are supposed to be conserved
\begin{align}
	\dot{\rho}_m &+ 3H\rho_m = 0,\nonumber \\
	\dot{\rho}_r &+ 4H\rho_r = 0, \nonumber \\
	\dot{\rho}_{de} &+ 3H(1 + w_{de})\rho_{de} = 0.
\end{align}

The equation of state (eos) parameter of DE $w_{de}$, should be defined in such a way that besides satisfying cosmological observations, it remains finite at both early and late times. A well-known example of such a parametrization was suggested by Chevallier, Polarski and Linder (CPL) and is defined as \cite{CPL}
\begin{align}
	w_{de}=w_0 + w_a \frac{z}{1+z},
\end{align}
where $w_0$ and $w_a$ are constants and $z$ is the redshift parameter defined as
\begin{align}
	1+z = \frac1a.
\end{align}
Beside observational successes of the CPL parameterization \cite{CPLexamples} the far future behavior of this parameterization is not well-defined. As one can easily check, the eos parameter of CPL diverges for $z\rightarrow-1$. 

A recent example of a parameterization which stays finite at early and late times and also at far future was introduced by Barboza and Alcaniz \cite{BA} which we will call it the BA parameterization of DE and is defined as
\begin{align}
	w_{de}(z) = w_0+w_a\frac{z(1+z)}{1+z^2},
\end{align}
where $w_0$ and $w_a$ are two constants which should be fixed from the cosmological observations. One can easily check that the above parameterization has the property
\begin{align}
	\lim_{z\rightarrow\infty,0,-1}w_{de}\rightarrow\textmd{finite}.
\end{align}
Although the BA parameterization of DE remains finite for all times, the behavior of the present time and the far future is the same since
\begin{align}
	\lim_{z\rightarrow0,-1}w_{de}=w_0.
\end{align}
This implies that the DE eos eventually loops back to its current value, regardless of its dynamics. This is an arbitrary constraint which does not occur for some famous DE models, like the scalar driven DE models \cite{scalarDE}. This implies that the BA kernel is non-monotonic and the future behavior is a duplicate of the near past. This is ad hoc and does not reflect the physical or observational ground. Also, the BA kernel has a fixed peak located at redshift $z\approx2.41$. As a result, the dark energy transition time is fixed in this model. 

In this paper, we will explore  a new extension of the BA (BAn) eos parameter of the form
\begin{align}
	w_{de} = w_0+\frac{n}{2}w_a\frac{(1+z)z^{n-1}}{1+z^n},
\end{align}
where $w_0$, $w_a$ and $n$ are model parameters. 
As one can see from the above proposal, the early, late and far future dynamics of the DE is finite, as in the BA model. However, there is a possibility that the dynamics of present and far future time differs from each other, depending on the value of $n$. For odd and integer values of parameter $n$, one can easily prove that
\begin{align}
	\lim_{z\rightarrow-1}w_{de}=w_0+\frac12 w_a,
\end{align}
which differs from the present value $w_{de}(z=0)=w_0$. Also, the location of the DE transition is now depends on $n$ and specially for odd values of $n$ the duplication behavior does not happen in the model in contrast to the original BA parametrization.

In this paper, we will not assume a priori value for the parameter $n$ and consider a 3D parametrization of the DE sector. It should be noted that BA2 is equivalent to the BA model.

As the inferred value of $n$ parameter is not necessarily an integer, in order to prevent unwanted imaginary values, we will assume that the eos parameter is written in the form
\begin{align}\label{omegade}
	w_{de} = w_0+\frac{n}{2}w_a\frac{(1+z)|z|^{n-1}}{1+|z|^n}.
\end{align}
From the conservation equation of the DE sector, one can obtain the DE energy density from
\begin{align}
	\rho_{de}=\rho_0\exp\left[3\int\frac{1+w_{de}}{1+z} dz\right],
\end{align}
with the result
\begin{align}\label{X}
	X(z)\equiv\frac{\rho_{de}}{\rho_0}=(1+z)^{3(1+w_0)}(1+|z|^n)^{\frac{3w_a}{2}}.
\end{align}
Now, defining the following set of dimensionless quantities
\begin{align}
\tau &= H_0 t, \quad H = H_0 h,\nonumber\\ \Omega_m(z) &= \frac{8\pi G}{3H^2}\rho_{m},\quad \Omega_{r}(z) = \frac{8\pi G}{3H^2}\rho_{r}(z),
\end{align}
one can obtain the dimensionless Hubble parameter as
\begin{align}\label{hub}
	h^2 = \Omega_{m0}(1+z)^3+\Omega_{r0}(1+z)^4+(1-\Omega_{m0}-\Omega_{r0})X,
\end{align}
where $\Omega_{i0}$ is the present time density abundances of the $i$th component.
\subsection{The impact of parameter $n$}
\begin{figure}
	\includegraphics[scale=0.45]{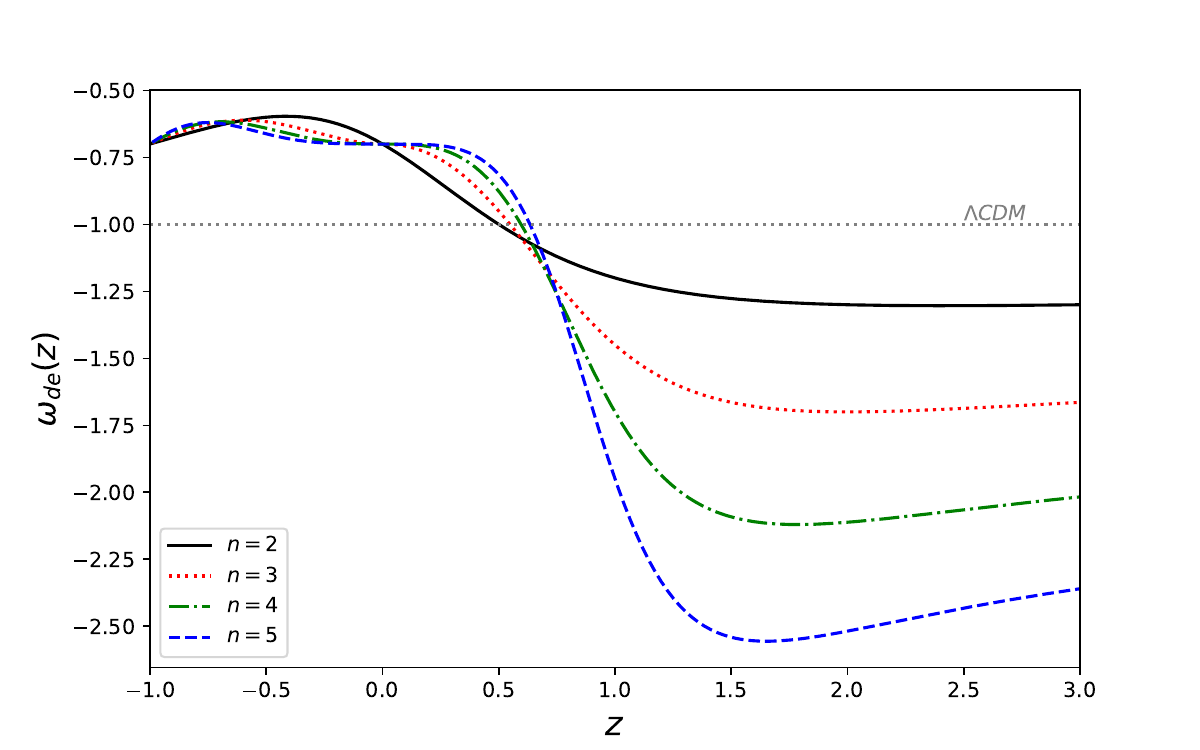}
	\caption{\label{figomegadecases} Evolution of the DE eos parameter 
		$w_{de}$ as a function of redshift for different values of $n=2$ 
		(solid), $n=3$ (dotted), $n=4$ (dash-dotted) and $n=5$ (dashed). For all curves, we used  $w_0=-0.7$, $w_a=-0.5$ and $\Omega_{m0}=0.3$.
		We have also indicated the $\Lambda$CDM model.}
\end{figure}
Before considering the full numerical analysis of the model~\eqref{hub}, 
it is instructive to investigate the impact of the new parameter $n$ on 
the cosmological dynamics of the Universe. We assume typical values for 
the model parameters as $w_0=-0.7$, $w_a=-0.5$ and $\Omega_{m0}=0.3$. 
In figure~\eqref{figomegadecases}, we have plotted the variation of $w_{de}$ 
as a function of redshift $z$ for different values of the model parameter 
$n=2,3,4,5$. It should be noted that the case $n=1$ is trivial since it 
reduces to a constant, i.e. the $\omega$CDM model. Values 
of $w_{de}$ above the $\Lambda$CDM value indicate quintessence behavior, 
while values below the $w=-1$ line represent phantom behavior. One can 
see from the figure that the qualitative behavior of all the models is 
the same, starting from the phantom regime at early times and making a 
phantom to quintessence crossing at redshift $z\in(0.5,1)$, transforming 
to quintessence at late times. However, for larger values of $n$, we 
observe a stronger phantom behavior and earlier phantom to quintessence 
crossing.
\begin{figure}[h]
	\includegraphics[scale=0.45]{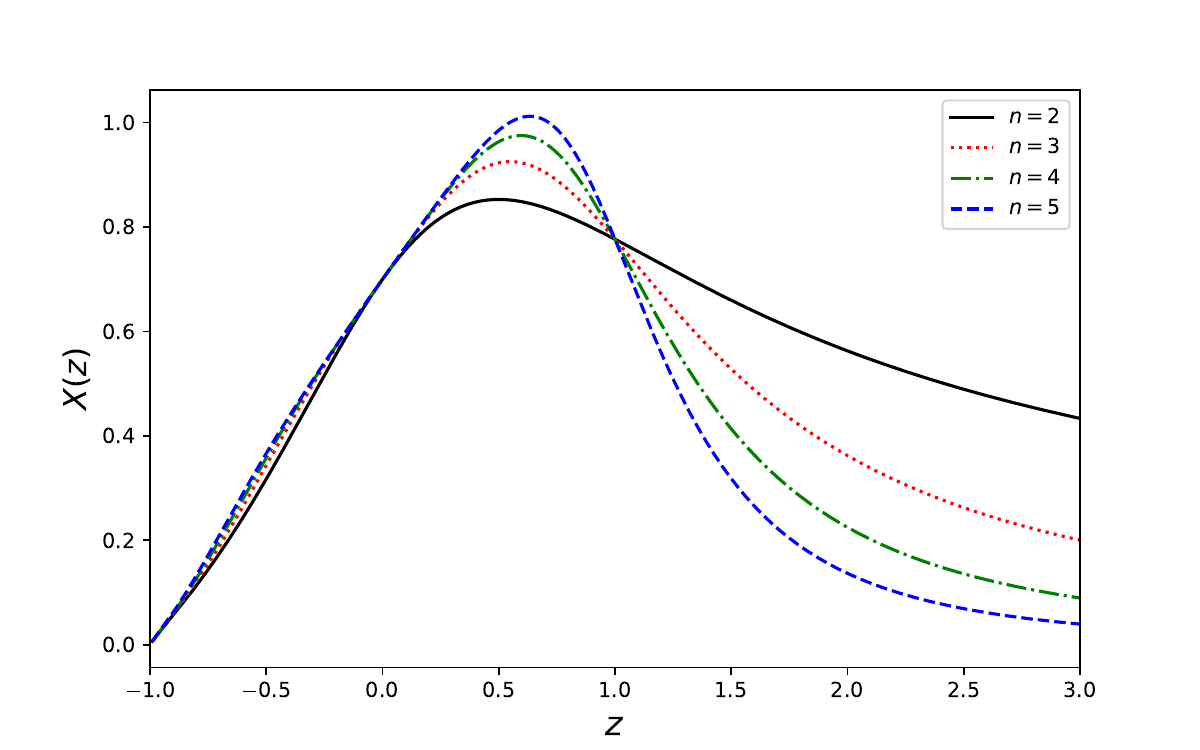}
	\caption{\label{figOmegadecases} Evolution of the rescaled DE energy 
		density $X$ as a function of redshift for different values of $n=2$ 
		(solid), $n=3$ (dotted), $n=4$ (dash-dotted) and $n=5$ (dashed). In all cases, we used  $w_0=-0.7$, $w_a=-0.5$ and $\Omega_{m0}=0.3$.}
\end{figure}

In figure~\eqref{figOmegadecases}, we have plotted the evolution of the 
rescaled DE energy density $X$ defined in~\eqref{X} as a function of 
redshift $z$ for different values of $n=2,3,4,5$. One can see from the 
figure that the DE energy density increases with time, reaches a maximum 
at around $z\in(0.5,1)$, and then decreases to its present day value. 
It can also be seen that larger $n$ results in a higher maximum value 
at earlier times.
\begin{figure}[h]
	\includegraphics[scale=0.45]{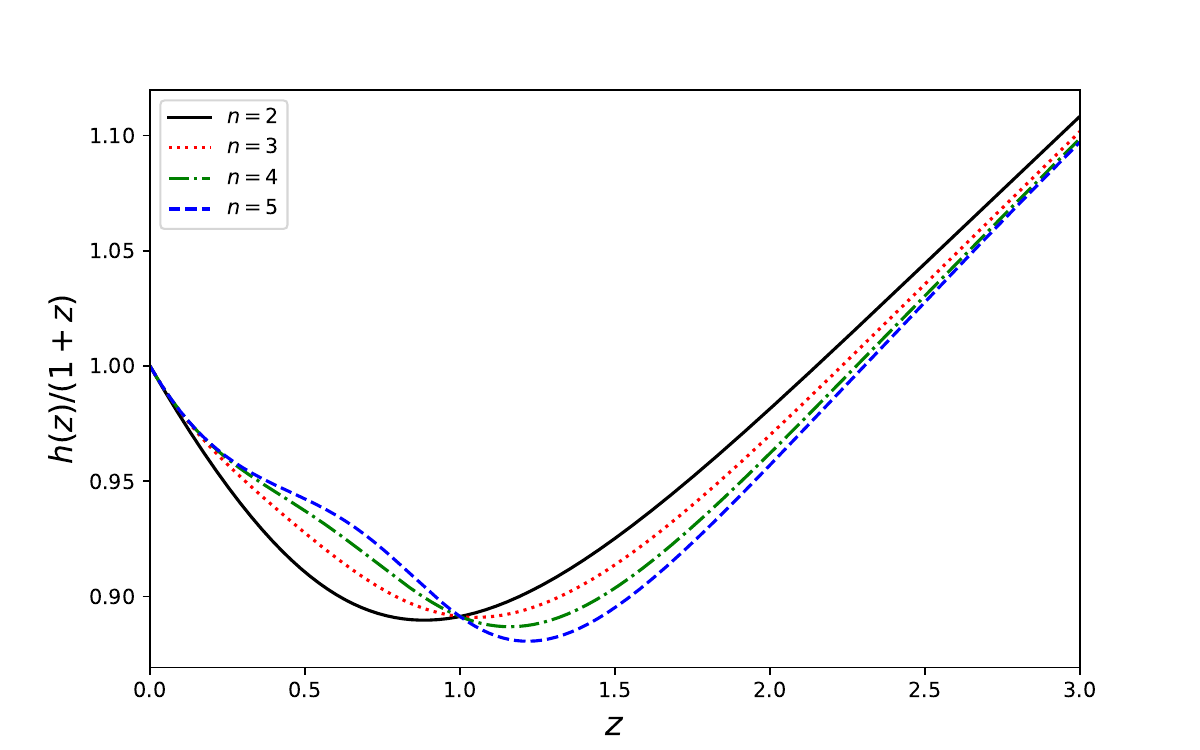}
	\caption{\label{fighubbleecases} Evolution of the rescaled Hubble 
		function $h/(1+z)$ as a function of redshift for different values 
		of $n=2$ (solid), $n=3$ (dotted), $n=4$ (dash-dotted) and $n=5$ 
		(dashed). For all curves, we used  $w_0=-0.7$, $w_a=-0.5$ and $\Omega_{m0}=0.3$.}
\end{figure}

For completeness, in figure~\eqref{fighubbleecases} we have plotted the 
evolution of the rescaled Hubble parameter $h(z)/(1+z)$ as a function 
of redshift $z$. Here again, one can see that the qualitative behavior 
for different values of $n$ is the same. However, for higher values of 
$n$, the minimum of the function occurs at a larger redshift, which 
implies that the deceleration to acceleration phase transition occurs 
at earlier times.

In summary, the qualitative behavior of the model is the same for 
different values of the parameter $n$, and a specific value of $n$ 
determines the details. In the following sections, we will perform a 
statistical analysis to infer the best value of $n$ based on 
cosmological observational data.

\section{Possible Lagrangian description} \label{sec3}
As we have mentioned before, the common feature of all the dynamical parameterizations of DE is that the cosmological equations are not obtained from an action principle. There are actually ways to write an action principle from scalar tensor theories but they are complicated and lacks physical interests \cite{scalardynamicalDE}. However, as was proved in \cite{farahzad}, these dynamical DE parameterizations can naturally be obtained from an action of specific models with non-standard matter Lagrangian. In this section we will summarize the main results of these type of theories and find a suitable form for the non-standard matter Lagrangian such that the BAn model arises.

Let us first introduce an action functional of the form
\begin{align}\label{act}
	S = \int d^4x \sqrt{-g} \left(\frac{1}{16\pi G}\, R + \mathcal{L}_m\right),
\end{align}
with the matter Lagrangian
\begin{align}\label{matterlag}
	\mathcal{L}_m=f(\rho,P),
\end{align}
where $\rho$ is the energy density and $P$ is the thermodynamics pressure of the baryonic matter which we will assume to be of a perfect fluid type. The special cases $f=-\rho$ and $f=P$ reduce to the standard Einstein general relativity with ordinary matter sources. 
The effective energy-momentum tensor is defined as
\begin{align}
	T^{\text{eff}}_{\mu\nu}=-\frac{2}{\sqrt{-g}}\frac{\delta(\sqrt{-g}\mathcal{L}_m)}{\delta g^{\mu\nu}}.
\end{align}
By varying the action \eqref{act} wrt to the metric tensor, one can obtain the Einstein field equation as
\begin{align}\label{eom1}
	G_{\mu \nu} = 8\pi G\, T^{\text{eff}}_{\mu \nu},
\end{align}
where $G_{\mu\nu}$ is the Einstein tensor. To obtain the effective energy-momentum tensor, one should determine the variation of thermodynamic quantities $\rho$ and $P$ wrt the metric. To do so, let us define the particle number flux and the Taub current as
\begin{align}
	J^\mu=\sqrt{-g}nu^\mu,\quad V_\mu=\mu^\prime u_\mu,
\end{align}
where $g$ is the metric determinant, $n$ is the particle number density which can be obtained as
\begin{align}
	n=\sqrt{\frac{g_{\mu \nu }J^{\mu }J^{\nu }}{g}},
\end{align}
$\mu^\prime$ is the enthalpy and $u^\mu$ is the 4-velocity of the matter fluid with condition $u_\mu u^\mu = -1$. 
The particle number flux and the Taub current, together with the entropy per particle $s$ are supposed to be independent of the metric tensor \cite{brown,secondvariation}
\begin{align}\label{assu1}
	\frac{ \delta s}{\delta g^{\alpha\beta}}=0,\quad
	\frac{\delta J^{\mu }}{\delta g^{\alpha\beta}}=0,\quad
	\frac{\delta V_\mu}{\delta g^{\alpha\beta}}=0.
\end{align}
As a result, with the help of first law of thermodynamics, one can find the variations of energy density and pressure as \cite{secondvariation}
\begin{align}
	\frac{\delta\rho}{\delta g^{\mu\nu}}&=\frac12(\rho+P)(g_{\mu\nu}+u_\mu u_\nu),\label{18-1}\\
	\frac{\delta P}{\delta g^{\mu\nu}}&=-\frac12(\rho+P)u_\mu u_\nu.\label{18-2}
\end{align}
With the help of above equations, the effective energy-momentum tensor can be obtained as
\begin{align}\label{emtensor1}
	T^{\text{eff}}_{\mu \nu} = (f_P - f_\rho) \, (\rho + P) \, u_\mu \, u_\nu + (f - f_\rho \, (\rho + P)) \, g_{\mu \nu},
\end{align}
where subscripts denote differentiation with respect to the argument.

Specific forms of the function $f$ can reproduce various parameterizations of DE from non-standard interactions of the baryonic matter sector. In our case, by defining the function $f$ to be
\begin{align}\label{lagcos}
	f(\rho,P)= P-B(\rho),
\end{align}
the cosmological field equations is reduced to
\begin{align}
	3H^2 &= 8\pi G(\, \rho +B),\label{frid1}\\
	3H^2 + \dot{H} &= -8\pi G\big(P -B + (\rho+P)B_\rho\big).\label{frid2}
\end{align}
The BA and BAn parameterizations of DE can then be realized by identifying the function $B$ to be equal to the energy density of the BA and BAn models. This can be done by denoting
\begin{align}
\rho = \rho_0 (1+z)^3,
\end{align}
which is obtained from the conservation equation of the baryonic sector. The result is
\begin{align}
	B(r)=r^{1+w_0}\left(1+(r^{\frac13}-1)^n\right)^\frac{3w_a}{2},
\end{align}
for the BAn model and we have defined $r\equiv\rho/\rho_0$. As we have discussed before, the BA model is a special case of the BAn model with $n=2$.

Despite that the above procedure could realize the DE parameterizations, it should be noted that because we have a non-standard matter Lagrangian, the conservation of the DE sector only holds in background FRW Universe where $\rho$ depends on the cosmic time. At perturbative level, the DE sector is not conserved and we have a possible interactions between dark matter and dark energy which means that the aformentioned procedure is different from the original DE parametrization.
\section{Statistical analysis} \label{sec4}
In order to constrain the model and cosmological parameters $H_0$, $\Omega_{m0}$, $w_0$, $w_a$ and $n$, we will use different combinations of the following datasets:
\subsubsection{Cosmic Chronometers}
The cosmic Chronometers (CC) is a direct and model-independent method for determining the Hubble parameter $H(z)$ by measuring the differential age evolution of passively evolving, massive early-type galaxies. Writing the Hubble parameter as
\begin{equation}
	H(z) = -\frac{1}{1+z}\frac{dz}{dt},
\end{equation}
the value of Hubble parameter $H(z)$, can be inferred from the measurement of age difference between two nearby galaxies separated by a small redshift interval $\Delta z$. In this paper we employ the 31 CC data points \cite{cc} which assumed to be independent.
The contribution of CC dataset to the total likelihood is
\begin{align}
	\chi^{2}_{\text{CC}} = \sum_i \left( \frac{H_{\text{obs},i} - H_{\text{th},i}}{\sigma_i} \right)^2,
\end{align}
where $i$ labels the data points, $H_{\text{obs},i}$ are the observational estimates of the Hubble parameter reconstructed from differential ages, $H_{\text{th},i}$ are the theoretical predictions of the model at corresponding redshifts, and $\sigma_i$ denotes the reported 1$\sigma$ uncertainties.

\subsubsection{Pantheon$^+$}
The Pantheon$^+$ compilation \cite{PANdata} represents the most updated and homogeneous collection of Type Ia supernova (SN Ia) distance measurements, consists of about 1500 spectroscopically confirmed SNe~Ia spanning the redshift range $0.001 < z < 2.26$, combining observations from 18 different surveys. Pantheon$^+$ improves the Pantheon dataset by enhancing the photometric calibration and refining light-curve fitting.
Here we employ the Pantheon$^+$ dataset without the SH0ES Cepheid calibration \cite{SH0ES}, so that the absolute magnitude $M$ is assumed to be a free parameter and will be inferred from the fitting.

The Pantheon$^+$ measurements are not independent and the covariance matrix is provided in \cite{PANdata}. 
The contribution of the Pantheon$^+$ dataset to the total likelihood is
\begin{equation}
	\chi^2_{\mathrm{Pantheon}^+}
	=
	\left[ \vec{\mu}_{\mathrm{obs}} - \vec{\mu}_{\mathrm{th}} \right]^{T}
	C^{-1}
	\left[ \vec{\mu}_{\mathrm{obs}} - \vec{\mu}_{\mathrm{th}} \right],
\end{equation}
where $C$ is the covariance matrix of the Pantheon$^+$ data.

\subsubsection{DESI(DR2) BAO}
We use the Baryon Acoustic Oscillation (BAO) measurements from the second data release of the Dark Energy Spectroscopic Instrument (DESI DR2) \cite{DESIDR2}. The dataset includes BAO observables extracted from several tracers of large-scale structure, namely the Bright Galaxy, the Luminous Red Galaxy and the Emission Line Galaxy samples and also quasars, covering the interval $0.2 \lesssim z \lesssim 2.4$. The DESI DR2 analysis reports BAO measurements in terms of distance ratios, typically of the form
\begin{equation}
	\frac{D_{\mathrm{M}}(z)}{r_{\mathrm{d}}}, \qquad
	\frac{D_{\mathrm{H}}(z)}{r_{\mathrm{d}}}, \qquad
	\frac{D_{\mathrm{V}}(z)}{r_{\mathrm{d}}},
\end{equation}
where $D_{\mathrm{M}}(z)$ is the comoving angular diameter distance, $D_{\mathrm{H}}(z)$ is the Hubble distance and $D_{\mathrm{V}}(z)$ is the spherically averaged distance. The quantity $r_{\mathrm{d}}$ denotes the sound horizon at the drag epoch which we will assume to be a free parameter and will infer from the fitting process. Defining the vector
\[
\vec{X}= 
\left(\frac{D_{\mathrm{M}}(z)}{r_{\mathrm{d}}},\; \frac{D_{\mathrm{H}}(z)}{r_{\mathrm{d}}},\; \frac{D_{\mathrm{V}}(z)}{r_{\mathrm{d}}} \right),
\]
the BAO contribution to the likelihood is then given by the $\chi^2$ function as
\begin{equation}
	\chi^{2}_{\mathrm{BAO}}
	=
	\left[ \vec{X}_{\mathrm{obs}} - \vec{X}_{\mathrm{th}} \right]^{T}
	C^{-1}
	\left[ \vec{X}_{\mathrm{obs}} - \vec{X}_{\mathrm{th}} \right],
\end{equation}
where $C$ is the covariance matrix.

\subsubsection{CMB Distance Priors}
To add an information from Cosmic Microwave Background (CMB)
without performing a full likelihood analysis, we adopt the CMB distance
priors extracted from the Planck observations \cite{Planck2018}. These
priors encode the geometric information of the CMB in a compressed form
through three quantities, the shift parameter $\mathcal{R}$, the angular
scale of the sound horizon at recombination $\ell_a$ and the baryon
density parameter $\Omega_b h^2$ which are defined as
\begin{equation}
	\mathcal{R} = \frac{\sqrt{\Omega_m H_0^2}}{c}\,D_{\mathrm{M}}(z_*),
	\qquad
	\ell_a = \frac{\pi\, D_{\mathrm{M}}(z_*)}{r_s},
\end{equation}
where $z_*$ is the redshift of photon decoupling and $r_s$ is the comoving sound horizon at $z_*$. Defining the
compressed data vector
\[
\vec{Y} = \left(\mathcal{R},\; \ell_a,\; \Omega_b h^2\right),
\]
the CMB contribution to the total likelihood is given by
\begin{equation}
	\chi^2_{\mathrm{CMB}}
	=
	\left[\vec{Y}_{\mathrm{obs}} - \vec{Y}_{\mathrm{th}}\right]^T
	C^{-1}
	\left[\vec{Y}_{\mathrm{obs}} - \vec{Y}_{\mathrm{th}}\right],
\end{equation}
where $C$ is the corresponding $3\times3$ covariance matrix provided
in \cite{Planck2018}.

\begin{table}[h]
	\centering
	\begin{tabular}{|c||c|c c c|}
		\hline
		Parameter & Prior & $\Lambda$CDM & BA & BAn \\
		\hline\hline
		$H_0$        & $\mathcal{U}(50, 80)$    
		& \checkmark & \checkmark & \checkmark \\
		$\Omega_{m0}$                     & $\mathcal{U}(0.2, 0.6)$ 
		& \checkmark & \checkmark & \checkmark \\
		$\mathcal{M}$                     & $\mathcal{U}(-30, -10)$  
		& \checkmark & \checkmark & \checkmark \\
		$\omega_0$                        & $\mathcal{U}(-2, 0)$     
		& $-$        & \checkmark & \checkmark \\
		$\omega_a$                        & $\mathcal{U}(-2, 0)$     
		& $-$        & \checkmark & \checkmark \\
		$n$                               & $\mathcal{U}(1, 3)$      
		& $-$        & $-$        & \checkmark \\
		$r_d$            & $\mathcal{U}(100, 200)$  
		& \checkmark & \checkmark & \checkmark \\
		$\Omega_b h^2$                    & $\mathcal{U}(0.01, 0.03)$
		& \checkmark & \checkmark & \checkmark \\
		\hline
	\end{tabular}
	\caption{Uniform priors on the free parameters for all models 
		and dataset combinations. A checkmark (\checkmark) indicates 
		that the parameter is varied in the corresponding model. $r_d$ and $\Omega_b h^2$ are only included when the BAO and CMB datasets are used, respectively. }
	\label{tabpriors}
\end{table}

\subsection{Dataset combinations}
In this paper, we will use three different combinations of the above datasets, namely\\
\begin{itemize}
	\item CC + Pantheon$^+$,
	\item CC + Pantheon$^+$ + BAO,
	\item CC + Pantheon$^+$ + BAO + CMB.
\end{itemize}
The likelihood function can then be obtained as
\begin{align}
	L=L_0e^{-\chi^2/2},
\end{align}
where $L_0$ is the normalization constant with corresponding loss function for each case. 
By maximizing the likelihood function, the best fit values of the parameters $H_0$, $\Omega_{m0}$,  $w_0$, $w_a$, $n$, $r_d$, $\mathcal{M}$ and $\Omega_bh^2$ can be obtained through the MCMC analysis. 
The Bayesian evidence was computed using the nested-sampling algorithm implemented in PolyChord \cite{pypolychord}. We used $nlive=500$ enabled clustering with precision criterion $\epsilon=0.01$, and adopted the uniform priors listed in table \ref{tabpriors}.

\begin{sidewaystable}
	\centering
	\caption{Constraints on cosmological parameters for $\Lambda$CDM, BA and BAn models from different data combinations. In the table CP, CPB and CPBC denote CC+Pantheon$^+$, CC+Pantheon$^+$+BAO and   CC+Pantheon$^+$+BAO+CMB respectively. The value of reduced $\chi^2_{red}$ is also shown for each dataset combinations.}
	\label{bestfit}
	
	\setlength{\tabcolsep}{3pt}
	\renewcommand{\arraystretch}{1.5}
	
	\begin{tabular}{|c||c|c|c||c|c|c||c|c|c|}
		\hline
		\multirow{2}{*}{Parameter} 
		& \multicolumn{3}{c||}{\textbf{$\Lambda$CDM}}
		& \multicolumn{3}{c||}{\textbf{BA}}
		& \multicolumn{3}{c|}{\textbf{BAn}} \\
		\cline{2-10}
		
		&\textbf{CP} & \textbf{CPB} &  \textbf{CPBC}
		& \textbf{CP} & \textbf{CPB} & \textbf{CPBC}
		& \textbf{CP} & \textbf{CPB} & \textbf{CPBC} \\
		\hline\hline
		
		$H_0$ & $66.549_{-1.6361}^{+1.6184}$ & $68.604_{-1.6542}^{+1.6244}$ & $70.405_{-0.4323}^{+0.4306}$
		& $66.656_{-1.6927}^{+1.7971}$ & $66.881_{-1.6181}^{+1.6241}$ & $69.378_{-0.6025}^{+0.5920}$
		& $66.670_{-1.6835}^{+1.7327}$ & $66.812_{-1.5835}^{+1.6102}$ & $69.304_{-0.6480}^{+0.6084}$ \\
		\hline
		$\Omega_b h^2$ & -- & -- & $0.022_{-0.0001}^{+0.0001}$ & -- & -- & $0.022_{-0.0002}^{+0.0001}$ & -- & -- & $0.022_{-0.0001}^{+0.0001}$ \\
		\hline
		$\Omega_{m0}$ & $0.357_{-0.0179}^{+0.0177}$ & $0.310_{-0.0081}^{+0.0079}$ & $0.313_{-0.0060}^{+0.0057}$
		& $0.361_{-0.0509}^{+0.0492}$ & $0.322_{-0.0125}^{+0.0131}$ & $0.324_{-0.0064}^{+0.0066}$
		& $0.349_{-0.0546}^{+0.0517}$ & $0.320_{-0.0128}^{+0.0128}$ & $0.324_{-0.0067}^{+0.0067}$ \\
		\hline
		$\mathcal{M}$ & $-19.451_{-0.0510}^{+0.0506}$ & $-19.400_{-0.0515}^{+0.0505}$ & $-19.343_{-0.0121}^{+0.0122}$
		& $-19.433_{-0.0547}^{+0.0550}$ & $-19.426_{-0.0509}^{+0.0509}$ & $-19.347_{-0.0154}^{+0.0156}$
		& $-19.431_{-0.0533}^{+0.0552}$ & $-19.429_{-0.0494}^{+0.0502}$ & $-19.348_{-0.0157}^{+0.0156}$ \\
		\hline
		$\omega_0$ & -- & -- & -- & $-0.799_{-0.0946}^{+0.0932}$ & $-0.792_{-0.0495}^{+0.0487}$ & $-0.800_{-0.0497}^{+0.0500}$
		& $-0.645_{-0.2198}^{+0.2518}$ & $-0.732_{-0.1137}^{+0.1299}$ & $-0.736_{-0.1127}^{+0.1353}$ \\
		\hline
		$\omega_a$ & -- & -- & -- & $-1.011_{-0.6657}^{+0.6733}$ & $-0.398_{-0.2001}^{+0.2061}$ & $-0.381_{-0.1156}^{+0.1145}$
		& $-1.096_{-0.6432}^{+0.6765}$ & $-0.486_{-0.2752}^{+0.2540}$ & $-0.500_{-0.2710}^{+0.2299}$ \\
		\hline
		$n$ & -- & -- & -- & -- & -- & -- & $1.733_{-0.4842}^{+0.5813}$ & $1.830_{-0.5388}^{+0.6476}$ & $1.865_{-0.5023}^{+0.6235}$ \\
		\hline
		$r_d$ & -- & $146.640_{-3.3590}^{+3.4743}$ & $142.475_{-0.4851}^{+0.4625}$ & -- & $147.093_{-3.3433}^{+3.4373}$ & $141.745_{-0.5118}^{+0.5245}$ & -- & $147.228_{-3.2925}^{+3.3187}$ & $141.816_{-0.5073}^{+0.5113}$ \\
		\hline\hline
		$\chi^2_{red}$ & $1.009_{-0.0012}^{+0.0011}$ & $1.015_{-0.0015}^{+0.0016}$ & $1.015_{-0.0017}^{+0.0016}$
		& $1.009_{-0.0016}^{+0.0016}$ & $1.007_{-0.0019}^{+0.0019}$ & $1.008_{-0.0020}^{+0.0021}$
		& $1.009_{-0.0018}^{+0.0018}$ & $1.008_{-0.0019}^{+0.0018}$ & $1.009_{-0.0021}^{+0.0021}$ \\
		\hline
	\end{tabular}
\end{sidewaystable}

In table \ref{bestfit}, we have summarized the best-fit values together with their $1\sigma$ uncertainties for the $\Lambda$CDM, BA, and BAn models. We have also reported the value of the reduced loss function $\chi^2_{red}$ defined as $$\chi^2_{red}=\frac{\chi^2}{\text{dof}},$$ where $\text{dof}$ is the number of degrees of freedom. It should be noted that the values of this quantity is around unity
for all models and combinations, meaning that the fit is statistically consistent with the observational uncertainties.

\subsection{Model comparisons}
\begin{figure*}
	\includegraphics[scale=0.6]{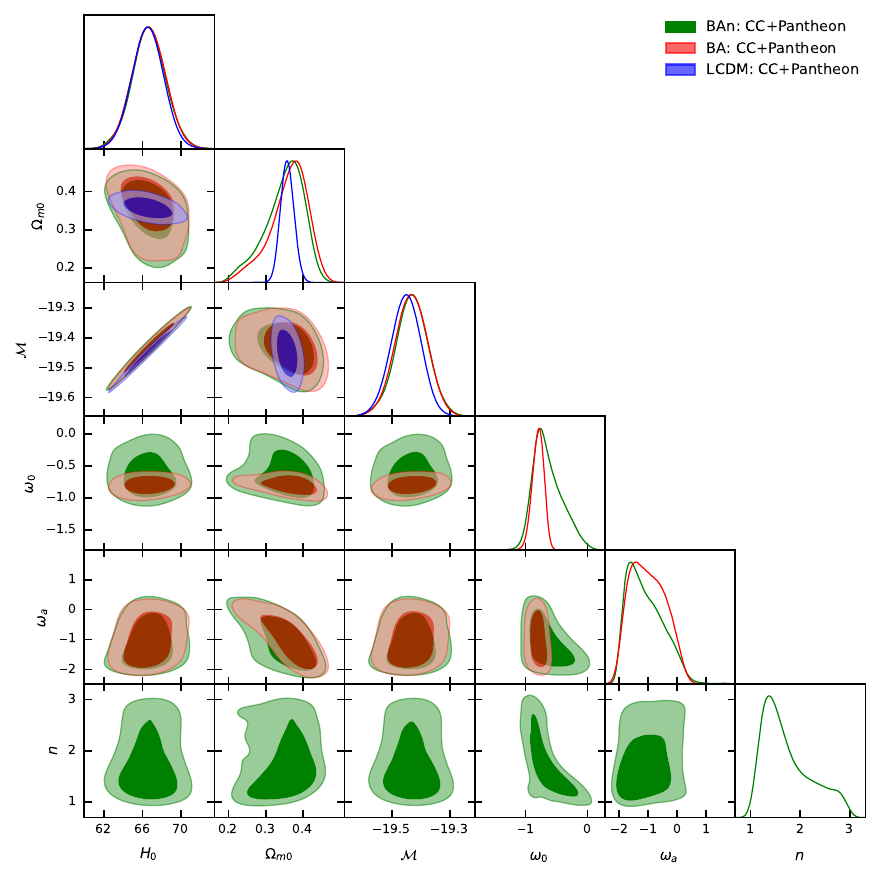}\includegraphics[scale=0.6]{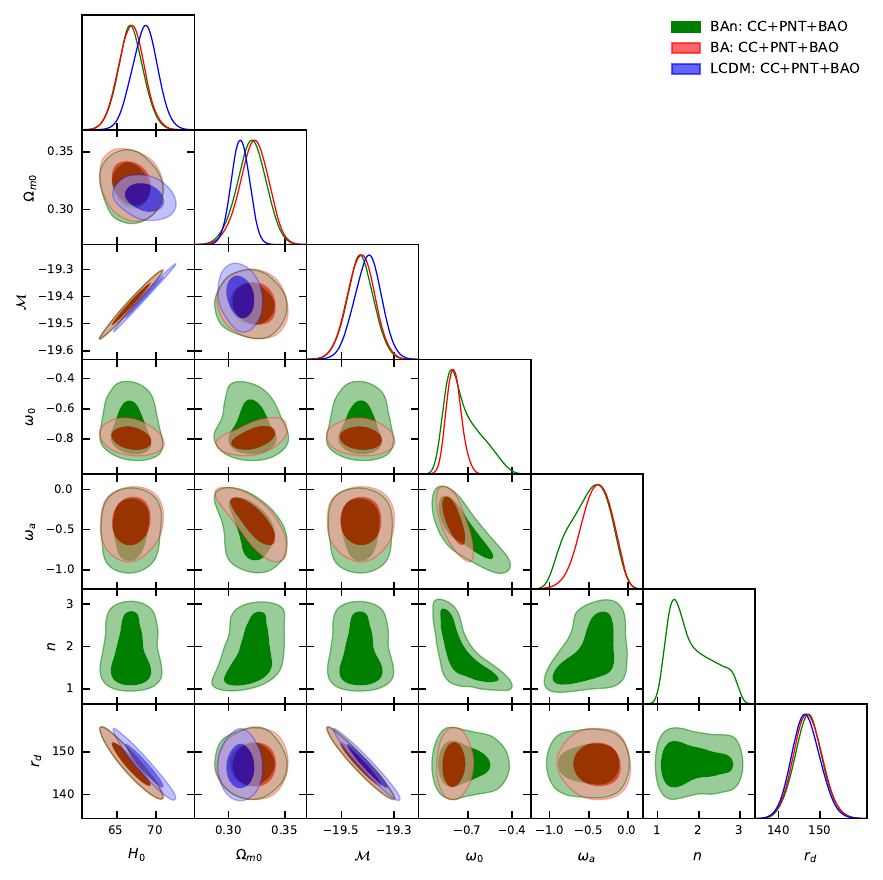}
	\includegraphics[scale=0.6]{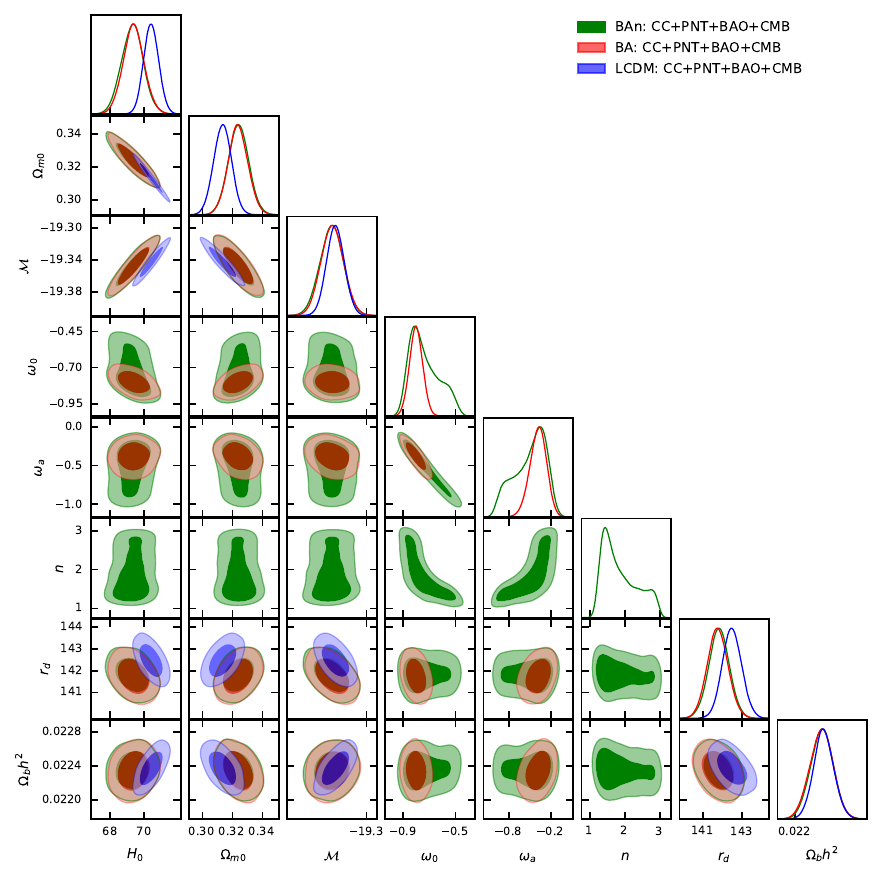}\includegraphics[scale=0.6]{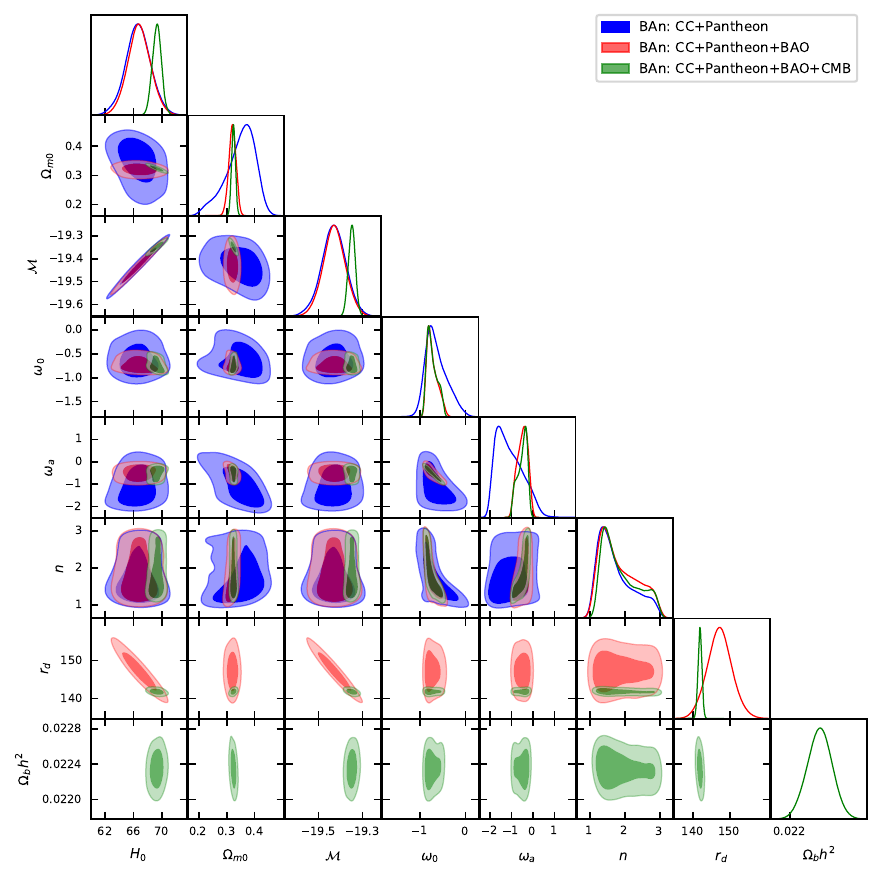}
	\caption{\label{corners} The corner plot of the values of parameters with their $1\sigma$ and $2\sigma$ confidence levels.}
\end{figure*}

In figure \eqref{corners} we have plotted corner plots for all the datasetset combinations and for all three $\Lambda$CDM, BA and BAn models. Also we have plotted the comparative corner plot of the BAn model for the three dataset combinations.

\begin{figure}
	\centering
	\includegraphics[scale=0.47]{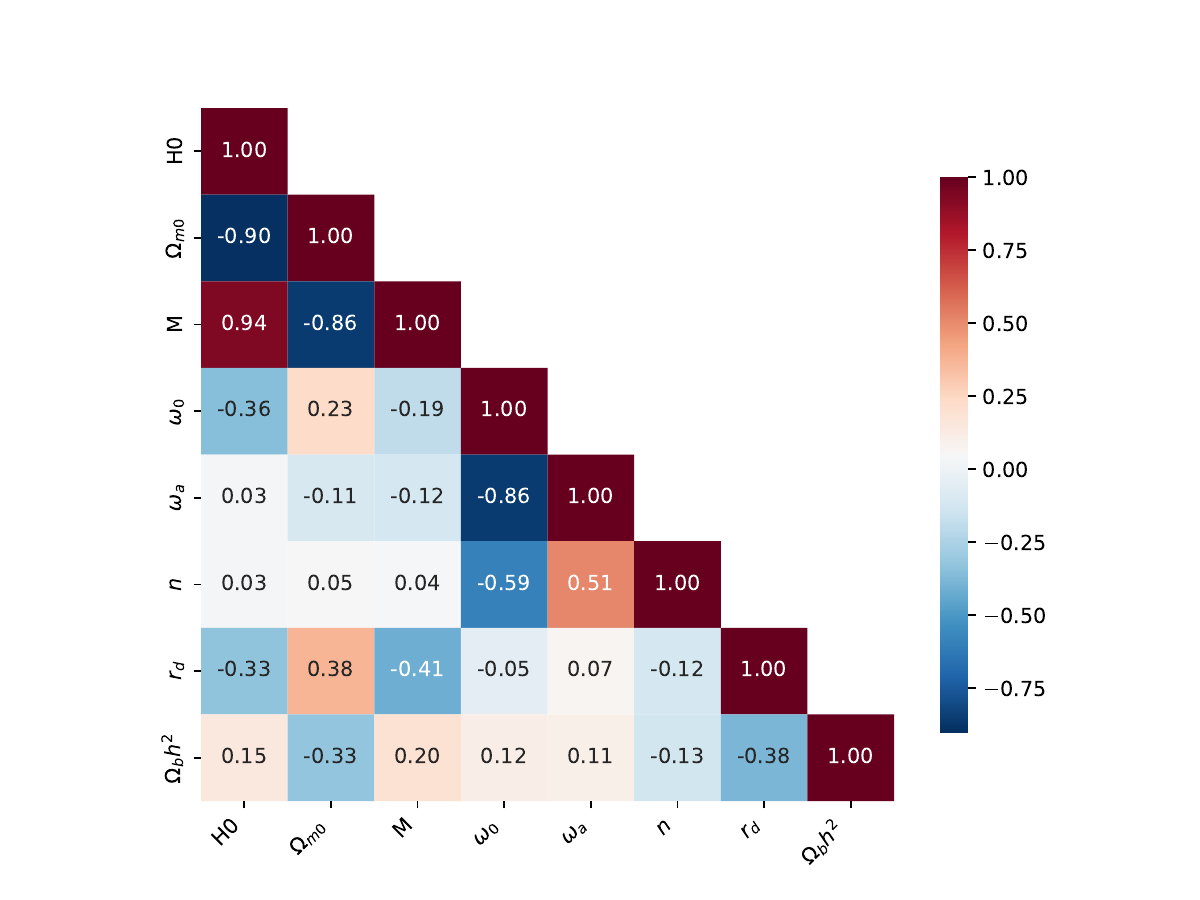}
	\caption{\label{pear} The Pearson correlation matrix between the parameters of  BAn model. Correlation of the model parameters $n$, $w_0$ and $w_a$ can be seen from the matrix.}
\end{figure}

From the corner plots, one can infer that besides standard parameter correlations between $H_0$, $\mathcal{M}$ and $r_d$ which is also present in the $\Lambda$CDM, model parameters $n$, $w_0$ and $w_a$ are also correlated. In figure \eqref{pear}, we have depicted the Pearson correlation matrix defined as
\begin{align}
	r_{ij} = \frac{\textmd{cov}(x_i,x_j)}{\sigma_i\sigma_j},
\end{align}
where $r_{ij}$ denotes the $ij$ component of the matrix obtained from the parameter pair $(x_i,x_j)$. Here $\textmd{cov}$ is the covariance and $\sigma$ is the standard deviation.
One can see that the parameters $w_0$ and  $w_a$ are strongly correlated. This also happens in the BA model and is not the specific feature of the present model. However, moderate correlation between the model parameter $n$ and the BA parameters $w_0$ and $w_a$ is new and shows that the parameter $n$ should not be fixed \textit{a priori} as we have in the BA model. This statement can be tested from the Bayes factor defined as
\begin{align}
	B_{AB} = \frac{\mathcal{Z}_{A}}{\mathcal{Z}_{B}},
\end{align}
where $\mathcal{Z}_i$ is the marginal likelihood of model $i$. The Bayes factor quantifies how strongly the data favors one model over the other. Here we adopt the Jeffreys scale \cite{jeffreys} in which $|\ln B_{AB}|<1$ indicate inconclusive evidence, $1<|\ln B_{AB}|<2.5$ indicate weak evidence, $2.5<|\ln B_{AB}|<5$ corresponds to moderate evidence and $|\ln B_{AB}|>5$ indicates strong evidence in favor of the model with higher evidence. With the above definition, larger positive Jefrreys scale indicate stronger evidence for model $B$.

\begin{table*}[ht]
	\centering
	\setlength{\tabcolsep}{3pt}
	\renewcommand{\arraystretch}{1.3}
	\begin{tabular}{|c||c|c|c|c|c||c|c|}
		\hline
		\textbf{Models} & \textbf{$\ln \mathcal{Z}$} & $\Delta \chi^2$ & $\Delta \mathrm{df}$ & \textbf{$p$-value} & $\ln B$ & \textbf{Jeffreys scale} & \textbf{Preferred by} $\begin{array}{c}\text{$\chi^2$} \\ \text{Evidence}\end{array}$ \\
		\hline\hline
		$\begin{array}{c}\textbf{BA} \\[-6pt]\text{vs }\\[-5pt] \Lambda\textbf{CDM}\end{array}$ & $\begin{array}{c}-905.286 \pm 0.262 \\-909.081 \pm 0.284\end{array}$ & $32.38$ & $1$ & $9.31 \times 10^{-8}$ & $3.795 \pm 0.386$ & $\begin{array}{c}\text{Moderate evidence}\end{array}$ & $\begin{array}{c}\text{Yes} \\ \text{Yes}\end{array}$ \\
		\hline
		$\begin{array}{c}\textbf{BAn} \\[-6pt] \text{vs }\\[-5pt] \Lambda\textbf{CDM}\end{array}$ & $\begin{array}{c}-903.801 \pm 0.229 \\ -909.081 \pm 0.284\end{array}$ & $33.92$ & $2$ & $4.31 \times 10^{-8}$ & $5.280 \pm 0.364$ & $\begin{array}{c}\text{Strong evidence}\end{array}$ & $\begin{array}{c}\text{Yes} \\ \text{Yes}\end{array}$ \\
		\hline
		$\begin{array}{c}\textbf{BAn} \\[-6pt] \text{vs}\\[-5pt] \textbf{BA}\end{array}$ & $\begin{array}{c}-903.801 \pm 0.229 \\ -905.286 \pm 0.262\end{array}$ & $1.54$ & $1$ & $2.15 \times 10^{-1}$ & $1.485 \pm 0.348$ & $\begin{array}{c}\text{Weak evidence}\end{array}$ & $\begin{array}{c}\text{No} \\ \text{Yes}\end{array}$ \\
		\hline
	\end{tabular}
	\caption{Model comparison for the largest dataset using Bayesian evidence and chi-square difference tests. The quantity $\Delta\chi^2$ is defined relative to the reference model and the p-value is computed under the nested-model assumption.}
	\label{bayes}
\end{table*}

In table \ref{bayes} we have presented the model comparison results based on both Bayesian evidence and the $\chi^2$ difference test for the CC + Pantheon$^+$ + BAO + CMB dataset. With respect to $\Lambda$CDM, both BA and BAn yield substantially improved fits, as reflected by the large values of $\Delta\chi^2$ and the corresponding very small $p$-values. In Bayesian terms, BA is moderately favored over $\Lambda$CDM, whereas BAn is strongly favored according to the Jeffreys scale.
On the other hand, between BAn and BA, the reduction in $\chi^2$ is modest and not statistically significant according to the $\chi^2$ difference test. Nevertheless, the Bayesian evidence shows weak support for BAn over BA. This suggests that the additional parameter in BAn may improve the overall model performance, although the current dataset does not provide decisive evidence in favor of the BAn model.
 
\section{cosmological implications} \label{sec5}
\begin{figure}
	\includegraphics[scale=0.45]{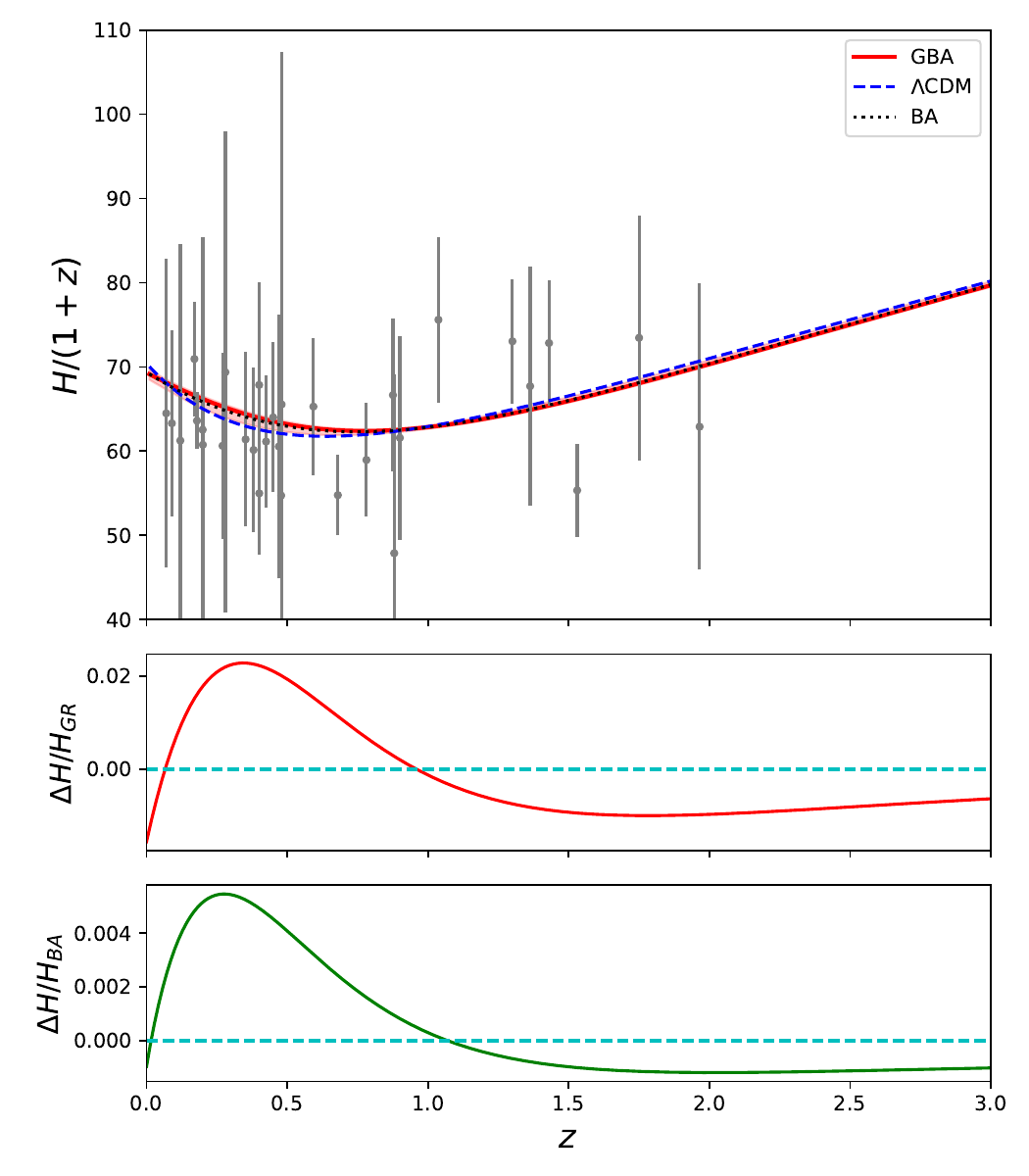}
	\caption{\label{fighubble}Evolution of the rescaled Hubble parameter $H/(1+z)$ as a function of redshift for  $\Lambda$CDM (dashed), BA (dotted) and BAn (solid) models. We have also depicted the relative difference between the three models. We have used the best fit parameters of the full dataset summarized in \ref{bestfit}. Shaded area denotes the 1$\sigma$ error for the BAn model.}
\end{figure}

Let us now consider cosmological implications of the BAn model. In figure \eqref{fighubble} we have plotted the evolution of the Hubble parameter for all three $\Lambda$CDM, BA and BAn models. We have also depicted the relative difference between the three models.
One can see from the figure that predictions of all the models are very similar. The difference between the BA and BAn models are very small and the maximum deviation is around $0.4\%$ which occurs at late times. The difference between BAn and $\Lambda$CDM models is a bit larger and reaches $2\%$ at late times.

\begin{figure}
	\includegraphics[scale=0.45]{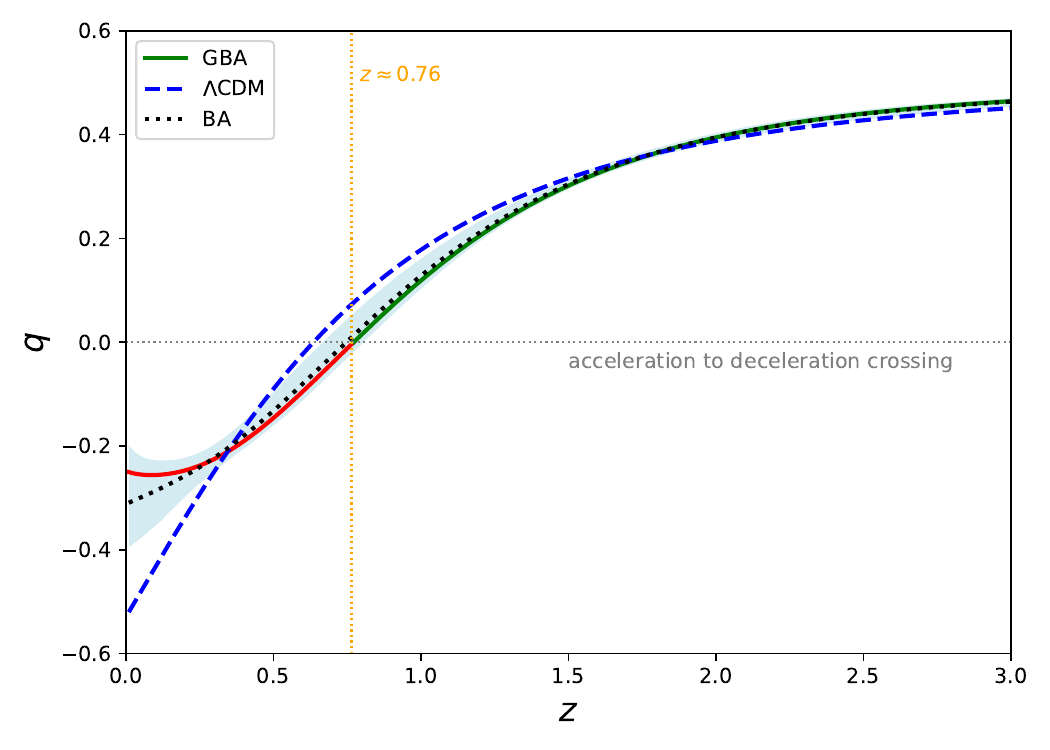}
	\caption{\label{figdec} Evolution of the deceleration parameter $q$ as a function of redshift for  $\Lambda$CDM (dashed), BA (dotted) and BAn (solid) models. We have also shown the exact deceleration to acceleration redshift for the BAn model. We have used the best fit parameters of the full dataset summarized in \ref{bestfit}. Shaded area denotes the 1$\sigma$ error for the BAn model.}
\end{figure}
\subsection{BAn cosmography}
The minimum of the rescaled Hubble diagram denotes the time of deceleration to acceleration phase transition. It is evident that the minimum for the $\Lambda$CDM model takes place at smaller redshifts compared to BA and BAn models. This indicates that the accelerated expanding era is younger in $\Lambda$CDM model.

In order to explore this, in figure \eqref{figdec} we have plotted the evolution of the deceleration parameter defined as
\begin{align}
 q = -1+(1+z)\frac{h^\prime}{h}.
\end{align}
One can see from the figure that the late time behavior of the models are significantly different with each  other. Although the $\Lambda$CDM model predicts more acceleration at the present time, the BA and BAn models predict slightly less acceleration. The acceleration of the BAn model deviates from the BA model at redshifts around 0.25, predicting even more less acceleration. This shows that the parameter $n$ in BAn model has more effects at late times, and deviations from the BA model are more significant there. On the other hand, the deceleration to acceleration phase transition redshift for the BAn model is $z\approx0.76$ which is more than the $\Lambda$CDM value $z\approx0.636$ and the BA value $z\approx0.747$. As we have mentioned earlier, this predicts younger acceleration period for $\Lambda$CDM model compared to the BAn and BA models.

Another feature of the Hubble diagram that can be seen from \eqref{fighubble} is that the slope of the BAn and BA curves are lower than their $\Lambda$CDM counterpart at late times. This can be seen more qualitatively from the jerk parameter defined as a third derivative of the scale factor. One can easily show that
\begin{figure}
	\includegraphics[scale=0.45]{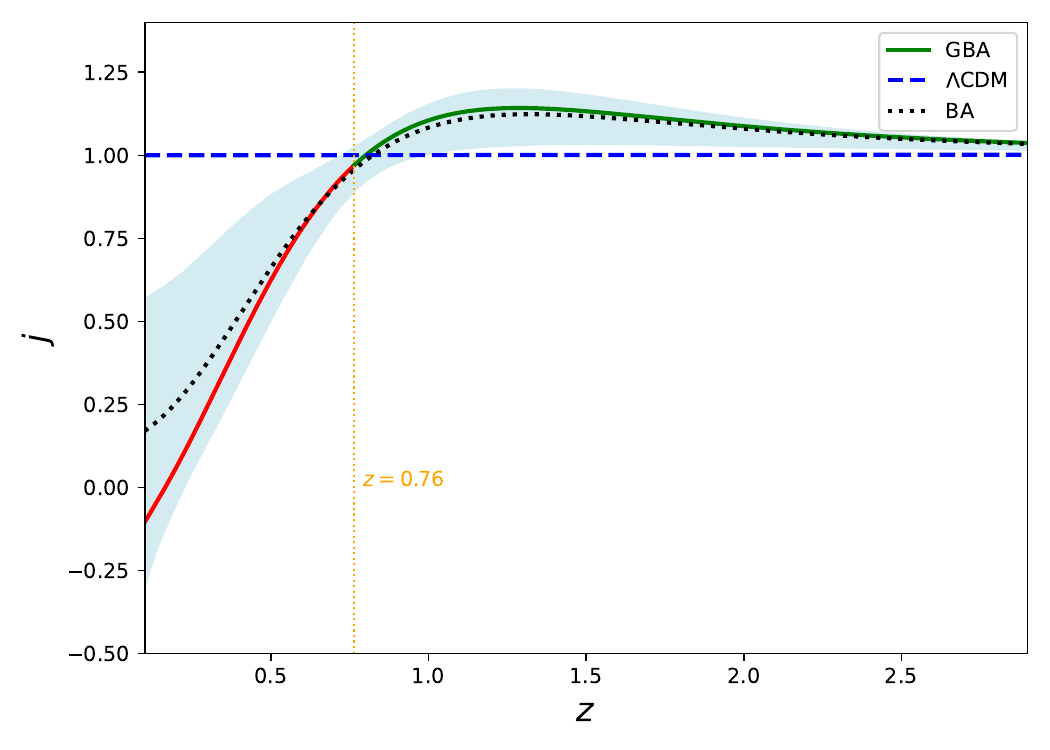}
	\caption{\label{figjerk} Evolution of the jerk parameter $j$ as a function of redshift for  $\Lambda$CDM (dashed), BA (dotted) and BAn (solid) models. We have also shown the exact deceleration to acceleration redshift for the BAn model. We have used the best fit parameters of the full dataset summarized in \ref{bestfit}. Shaded area denotes the 1$\sigma$ error for the BAn model.}
\end{figure}
\begin{align}
	j(z) = 2q^2 + q + (1 + z)\frac{dq}{dz}.
\end{align}
The jerk parameter could be seen as a characteristic of modified gravity theories since the value in $\Lambda$CDM model is exactly equal to unity. As a result, any deviations of the jerk parameter from unity is the characteristic of modified gravity. Positive/negative values of the jerk parameter represents higher/lower slope in the Hubble diagram. It can be seen from the jerk parameter that the slope of both BA and BAn models are higher at early times and becomes lower at late times in line with the predictions from the Hubble diagram \eqref{fighubble}. The $\Lambda$CDM crossing takes place at around the deceleration to acceleration crossing redshift. Also, it is evident that the highest slope at the present time belongs to the BAn model. 

\begin{figure}
	\includegraphics[scale=0.5]{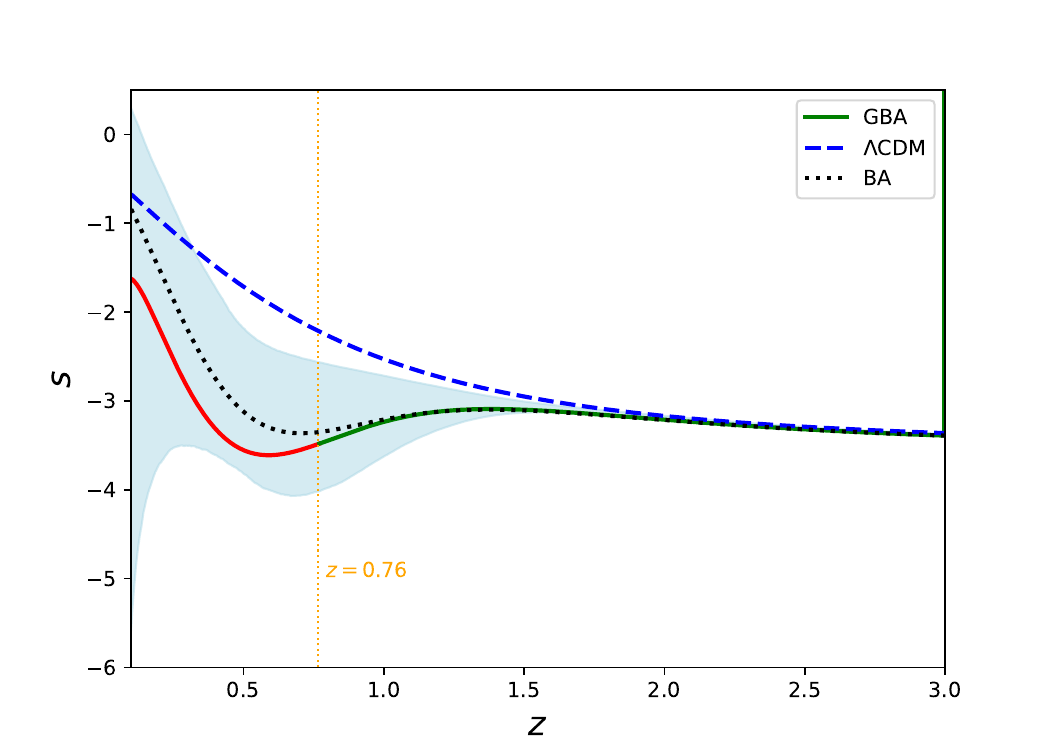}
	\caption{\label{figsnap} Evolution of the snap parameter $j$ as a function of redshift for  $\Lambda$CDM (dashed), BA (dotted) and BAn (solid) models. We have also shown the exact deceleration to acceleration redshift for the BAn model. We have used the best fit parameters of the full dataset summarized in \ref{bestfit}. Shaded area denotes the 1$\sigma$ error for the BAn model.}
\end{figure}

To complete the cosmography discussion of the BAn model, in figure \eqref{figsnap} we have plotted the evolution of the snap parameter defined as
\begin{align}
	s=-(1+z)\frac{dj}{dz}-2j-3jq.
\end{align}
It is evident from the figure that the snap parameter for both BA and BAn models lie below the $\Lambda$CDM curve. Also, the snap parameter of all models coincide at earlier times, and similar to the jerk parameter, the highest derivative at the present times belong to the BAn model.
 
 \subsection{The DE behavior}
\begin{figure}
 	\includegraphics[scale=0.47]{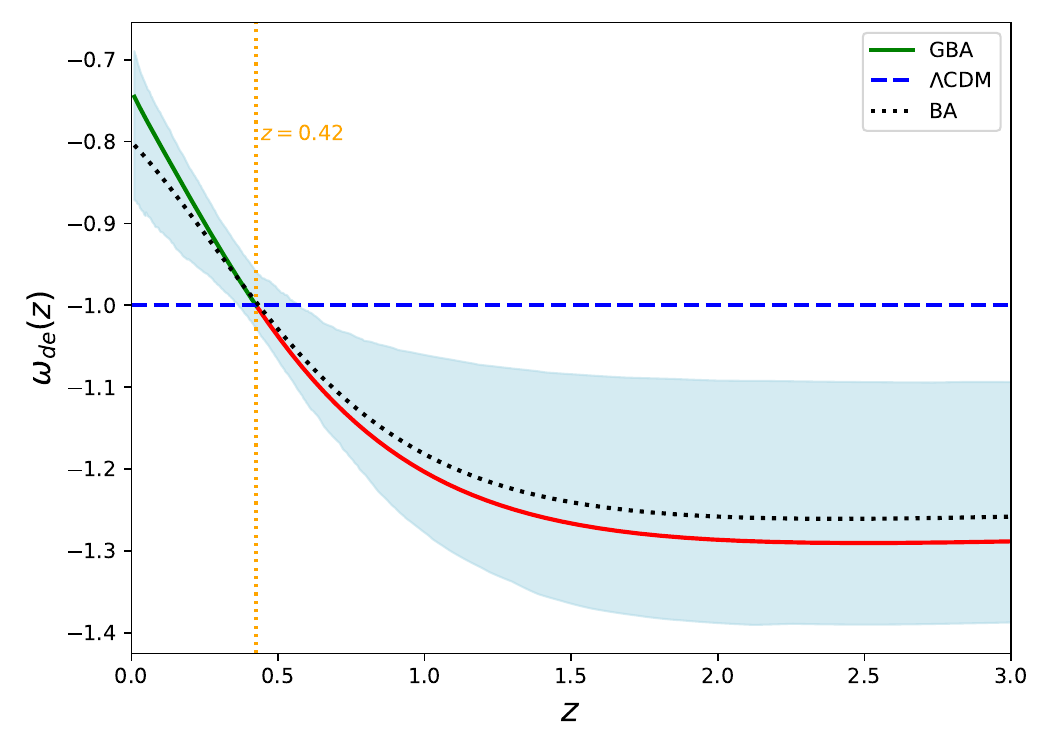}
 	\caption{\label{figomegade} Evolution of the DE eos parameter $w_{de}$ as a function of redshift for  $\Lambda$CDM (dashed), BA (dotted) and BAn (solid) models. We have also shown the exact quintessence to phantom transition redshift for the BAn model. We have used the best fit parameters of the full dataset summarized in \ref{bestfit}. Shaded area denotes the 1$\sigma$ error for the BAn model.}
 \end{figure}

 Let us now concentrate on the dark energy part of the model. In figure \eqref{figomegade}, we have plotted the behavior of the DE eos parameter $w_{de}$ defined in \eqref{omegade} as a function of the redshift.
 The DE eos parameter for the $\Lambda$CDM model is equal to $-1$. Values below $-1$ belongs to the phantom DE models and values in range $-1<w_{de}<-1/3$ describe quintessence DE models. It can be seen from the figure that both the BA and BAn models are phantom-like at earlier times while transitioning to quintessence behavior at late times. The crossing redshift for the BAn model is at $z\approx0.42$ which is approximately equal to the transition redshift of the BA model. However, the BAn model predicts stronger phantom behavior and weaker quintessence behavior compared to the BA model. Remembering that the deceleration to acceleration crossing redshift is at $z\approx0.76$, one can see that in the acceleration phase of the Universe, the DE is first phantom like and then make a transition to quintessence. 
 
\begin{figure}
 	\includegraphics[scale=0.5]{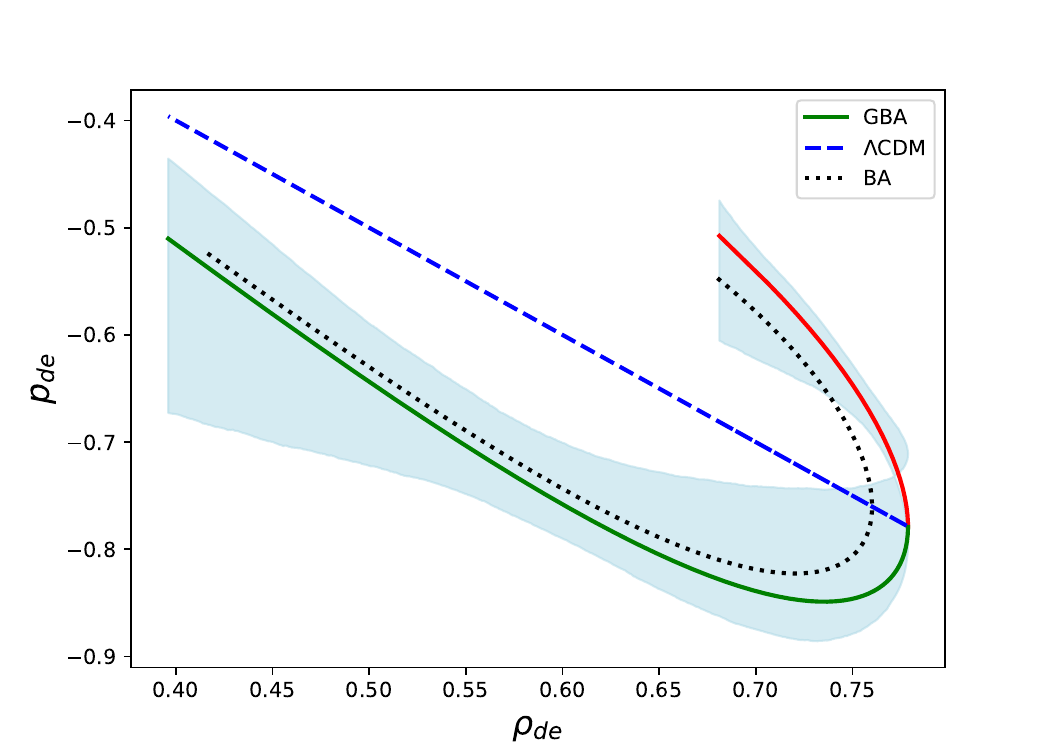}
 	\caption{\label{figpasrho} Evolution of the DE pressure $P_{de}$ as a function of DE energy density $\rho_{de}$ for  $\Lambda$CDM (dashed), BA (dotted) and BAn (solid) models. We have used the best fit parameters of the full dataset summarized in \ref{bestfit}. Shaded area denotes the 1$\sigma$ error for the BAn model.}
 \end{figure}

 For further investigations, in figure \eqref{figpasrho}, we have plotted the evolution of the DE pressure as a function of the DE energy density for all the three models.
 It can be seen that the behavior of the BA and BAn models are similar and both lies near the $\Lambda$CDM curve. It should be noted that the portion of the curves above/below the $\Lambda$CDM curve belongs to the phantom/quintessence regimes. The behavior of the models are in line with the predictions we had in figure \eqref{figomegade}. What can be inferred from this plot is that the BAn model make the DE evolution a bit broaden compared to the BA model.
    \begin{figure*}
 	\includegraphics[scale=0.34]{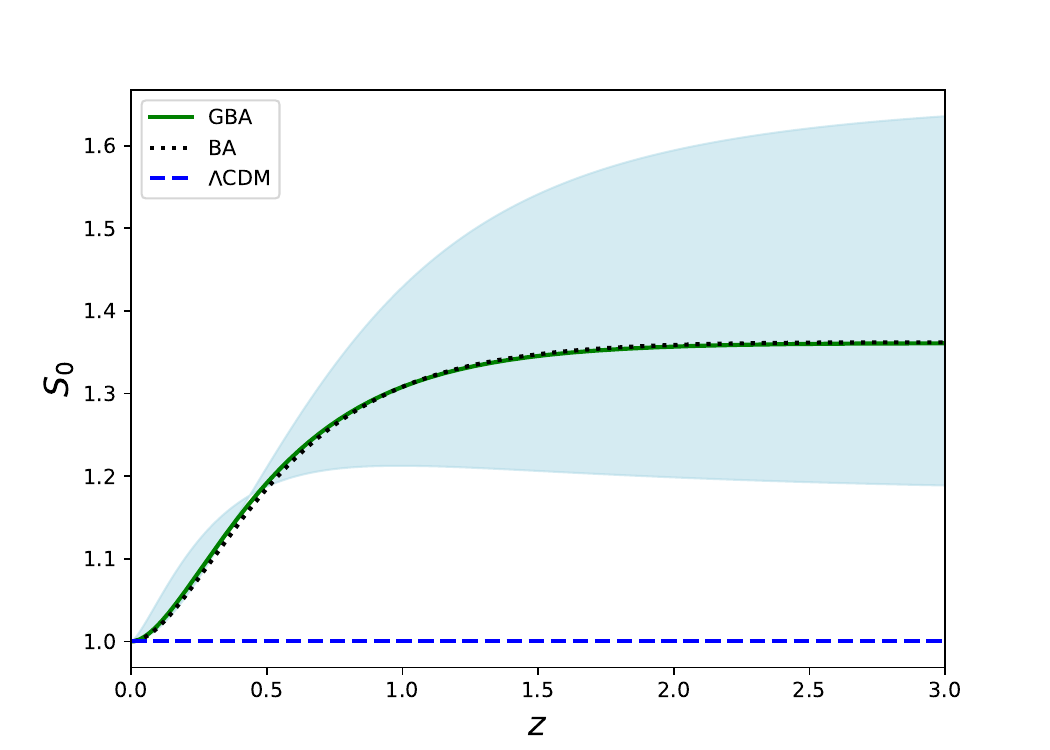}\includegraphics[scale=0.34]{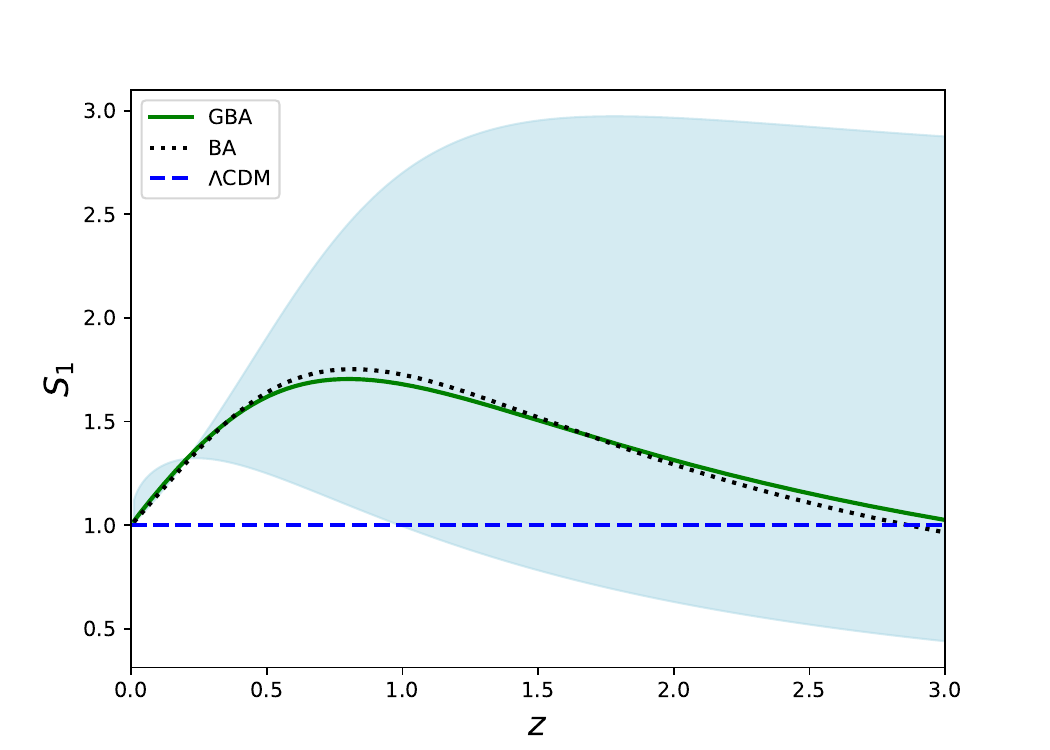}\includegraphics[scale=0.34]{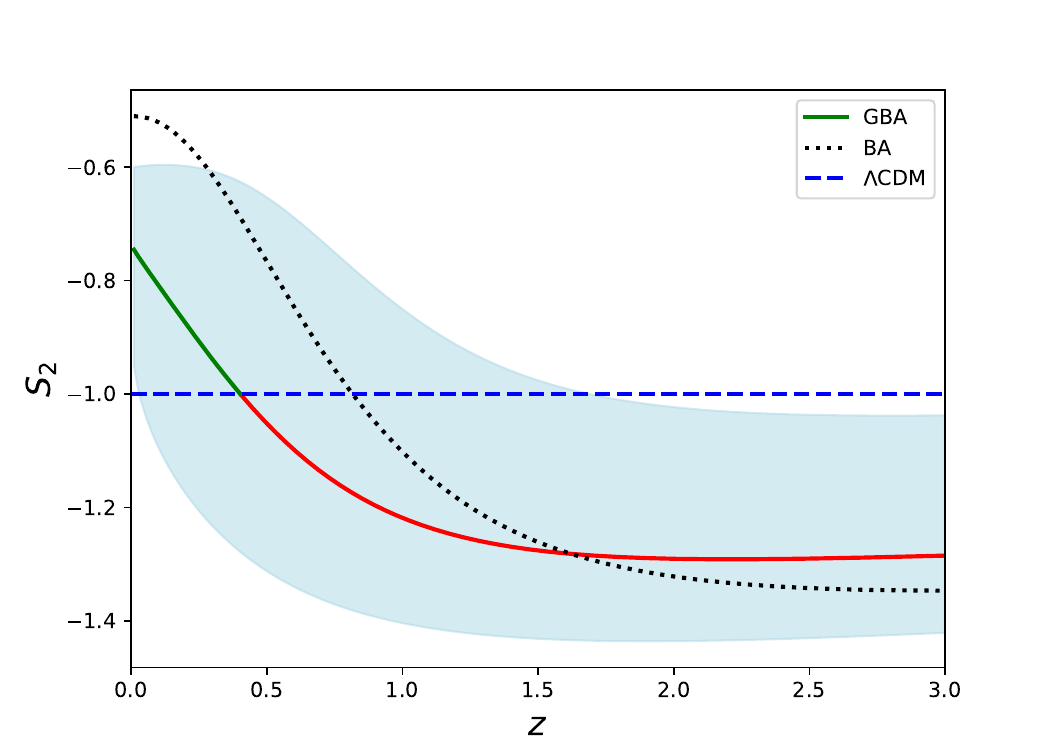}
 	\caption{\label{figshape} The behavior of the shape functions as a function of redshift $z$ for  $\Lambda$CDM (dashed), BA (dotted) and BAn (solid) models. We have used the best fit parameters of the full dataset summarized in \ref{bestfit}. Shaded area denotes the 1$\sigma$ error for the BAn model.}
 \end{figure*}
 
 In order to further investigate the DE behavior of the model, let us consider the newly proposed DE shape functions defined as \cite{shapepaper}
 \begin{align}
 	S_0(z) &= (1+z)^{-3}+\frac{X(z)(1+z)^{-3}-1}{w_{de}(0)},\nonumber\\
 	S_1(z) &=\frac{P_{de}(z)}{P_{de}(0)},\nonumber\\
 	S_2(z)& = w_{de}(z)+\frac13(1+z)\frac{w_{de}^\prime(z)}{w_{de}(z)},
 \end{align}
 where prime denotes derivative wrt the redshift $z$ and we have defined $X(z)=\rho_{de}(z)/\rho_{de}(0)$ as before. It should be noted that for $\Lambda$CDM model we obtain
 \begin{align}
 	S_0\rightarrow1,\quad S_1\rightarrow1,\quad S_2\rightarrow-1.
 \end{align}
The shape functions $S_1$ and $S_2$ are dimensionless energy density and pressure of the DE sector, shifted so that for $\Lambda$CDM model both will tend to unity. The $S_3$ function captures the behavior of the DE eos parameter.
 
 In figure \eqref{figshape}, we have plotted the shape functions for the BAn model with their 1$\sigma$ error. We have also denoted the behavior of the $\Lambda$CDM and BA models. As can  be seen from the figure,  the shape functions $S_0$ and $S_1$ for both BA and BAn models behave approximately the same. The functional behavior are also the same as other dynamical DE parameterizations \cite{shapepaper} which is the common characteristic feature of the DE models. The shape function $S_3$ captures the behavior of the eos parameter. It can be seen that the $\Lambda$CDM crossing occurs later in BAn model compared to the BA model. This is in line with our previous observations on the behavior the BAn model. Also, the slope of the BAn model is lower compared to the BA model signaling smoother evolution in time.
  \begin{figure}
	\includegraphics[scale=0.5]{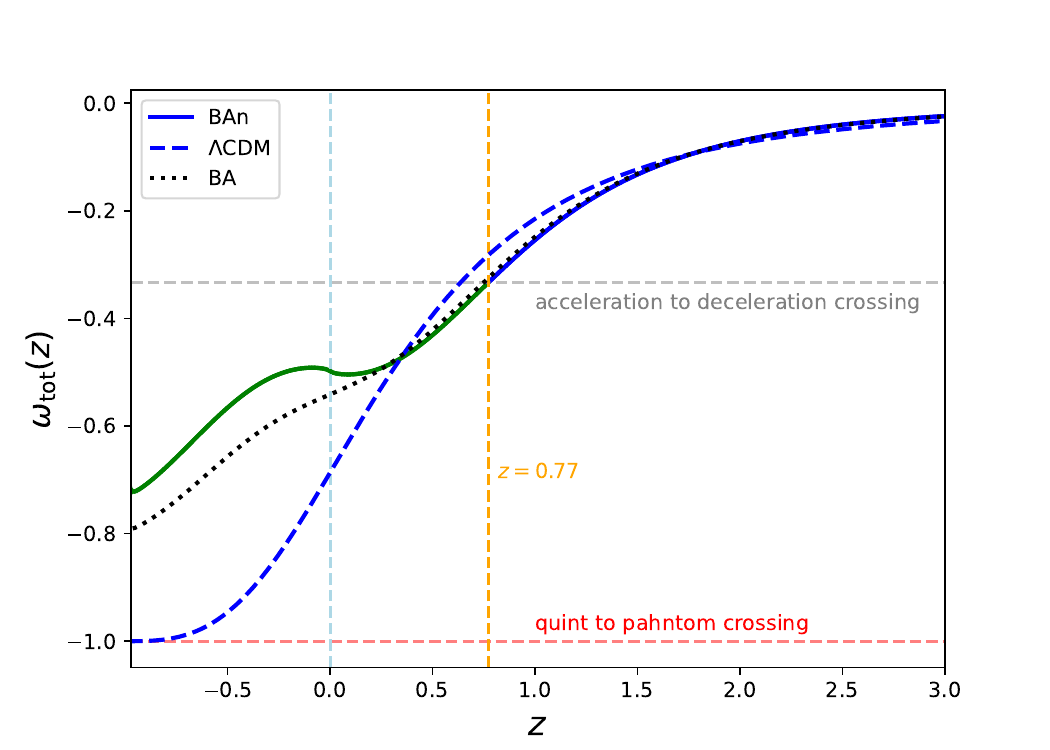}
	\caption{\label{figomegaeff} Evolution of the total eos  parameter $w_{tot}$ as a function of redshift $z$ for  $\Lambda$CDM (dashed), BA (dotted) and BAn (solid) models. We have also shown the exact deceleration to acceleration redshift for the BAn model. We have used the best fit parameters of the full dataset summarized in \ref{bestfit}.}
\end{figure}
\subsection{The total behavior}
In figure \eqref{figomegaeff} we have plotted the total eos parameter $w_{tot}$ defined as
\begin{align}
	w_{tot} = \frac{P_r+P_{de}}{\rho_m+\rho_r+\rho_{de}}.
\end{align}
The total eos parameter determines the behavior of the Universe as a whole. The negative range of the redshift corresponds to the future time, being $z\rightarrow-1$ equivalent to the future infinity. In the figure, we have also indicated the acceleration to deceleration crossing and  also the quintessence to phantom crossing lines. One can see from the figure that the Universe has a quintessence-like accelerated expansion for its entire lifetime after the deceleration to acceleration crossing. The same is also true for the BA model. It should be noted that this is in contrast to the $\Lambda$CDM model where the Universe reaches the de Sitter expansion rate at future infinity. Also, one can see again that the acceleration rate is smaller for the BAn model compared to the BA model. 

\section{Conclusion and final remarks}\label{sec6}
In this paper, we have considered a new generalization of the Barboza-Alcaniz parametrization of dark energy in which the future behavior is modified. The BA parametrization has two free parameters which controls the late and early behavior of the DE sector. However, the future behavior of the model at $z<0$ is a duplicate of the $z>0$ era which is artificial. Since the main cause of the above property is the power of the redshift in the denominator of the BA kernel, we have relaxed this power to an arbitrary parameter. As a result, the BAn parametrization of DE is a three-parameter model. We have confronted the new parametrization with four independent datasets, namely the Cosmic chronometers, the Pantheon$^+$ catalog, the BAO data from new DESI DR2 release and the CMB distance priors which reflect the geometric information of the full CMB dataset from Planck observations. One of the main results of the inference is that the new parameter $n$ is moderately correlated with the other DE parameters, namely $\omega_0$ and $\omega_a$. This indicates that the new parameter is observationally viable and that freeing it increases our information about the DE behavior. However, we have seen that the final value of the parameter $n\approx1.865$ is not far from the original choice, suggesting that the BA parameterization is also reliable. This could also be seen as a consistency check on the BA model. 

As we have shown in the paper, the introduction of this new $n$ parameter, make the BAn model the most favorable by cosmological observations compared to the original BA and also the $\Lambda$CDM models. This is not entirely surprising, since statistically speaking, a larger number of free parameters generally improves the fit to data. However, our new parameter seems to be necessary as we have discussed above.

The cosmography of this new parameterization indicates that the qualitative behavior of the BAn model is similar to the BA parametrization. However, the BAn model seems to be slightly more moderate. As we have seen in this paper, the BAn model produce less acceleration for the Universe at the present time and also in the future while the DE sector shows more phantom-like behavior at early times. This can also be seen in the evolution of $\omega_{tot}$ where both BA and BAn models end in the quintessence regime while the $\Lambda$CDM model tends to de Sitter expansion at future infinity. However, the BAn model has a smaller acceleration at future infinity, makes it milder compare to BA model. We have also plotted the shape functions of BAn  and BA models indicating the above feature directly in the behavior of $S_3$.

Finally, the new generalization of the BA model can be taken seriously as a means to shed more light on the behavior of DE.

\end{document}